%% file: main.tex
\documentclass[10pt, twocolumn,secnumarabic,amssymb, nobibnotes, floatfix, aps, prd]{revtex4-2}

\usepackage[utf8]{inputenc}
\usepackage{graphicx}
\graphicspath{{./}}
\usepackage{subfigure}
\usepackage{booktabs}
\usepackage{import}
\usepackage{hyperref}
\usepackage{amsmath}
\usepackage{color}
\usepackage{layouts}

\usepackage{enumitem}
\newenvironment{algorithm}[1][\arabic*.]
    {
        \newlist{enumtt}{enumerate}{1}
        \setlist[enumtt,1]{label={#1},before={\ttfamily}}

        \hfill \break
        \begin{minipage}{\linewidth}
        \begin{enumtt}
    }
    {
        \end{enumtt}
        \end{minipage} \\ \\
    }

\begin{document}

\title{Site-Specific Parameterization of Ocean Spectra for Power Estimates of Wave Energy Converters}%

\author{Rafael Baez Ramirez}
\email{rbaezra@sandia.gov}
\affiliation{Sandia National Laboratories, NM 87123}

\author{Ethan J. Sloan}
\email{esloan03@unm.edu}
\affiliation{University of New Mexico, NM 87106}

\author{Carlos A. Michelén~Ströfer}
\email{cmichel@sandia.gov}
\affiliation{Sandia National Laboratories, NM 87123}

\date{\today}

\begin{abstract}
    Estimating the mean annual power of a wave energy converter (WEC) through the method of bins relies on a parametric representation of all possible sea states. 
    In practice, two-parameter spectra based on significant wave height and energy period are ubiquitous. 
    Two-parameter spectra have been shown insufficient in capturing the range of spectral shapes that can occur in an actual ocean environment.
    Furthermore, through sensitivity analysis, these two-parameters have been shown to be insufficient for predicting power performance of WECs.
    Four parameter spectra, which expand the parameter space to include two additional \emph{shape parameters} have been shown sufficient in capturing sea state variance, but their effect on mean power estimates has not been presented. 
    This work directly looks at the effects of incorporating 4-parameter spectra into annual power estimates compared to using the traditional 2-parameter spectra. 
    We use two different 4-parameter spectra: one from the literature and a novel machine learning-based autoencoder, presented here.
    Both are shown to improve the information retained when parameterizing spectra.
    The site-specific autoencoder performs consistently the best across two case studies of mean annual power prediction, achieving an error around 1\% in each instance. 
    The 2-parameter spectra resulted in less consistent predictive performance, with errors of $-$8\% and 1\% in the two case studies. 
    For the case study where all three models performed well, it is shown that the low error in the 2-parameter model is attributable to a symmetrical distribution of large errors whereas both 4-parameter spectra result in relatively low errors throughout the parametric space.
    These results highlight the need for more sophisticated resource characterization methods for estimating the power performance of WECs and suggest site-specific machine learning-based spectra are an adequate option.
\end{abstract}

\maketitle

\begin{description} 
    \item[Keywords] wave energy converter, wave resource, wave spectra, machine learning, autoencoder, annual power estimate
\end{description}

\tableofcontents

\section{Introduction}
\begin{figure}
        \centering
        \subfigure[]{\includegraphics[width=0.95\linewidth]{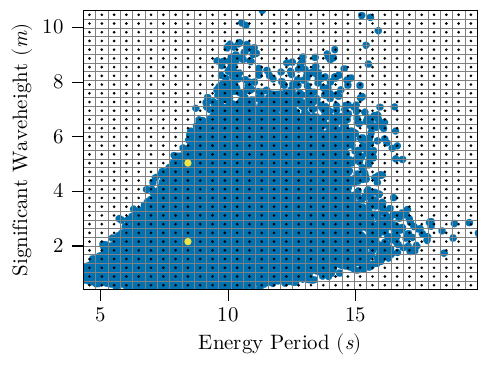}} \\
        \subfigure[]{\includegraphics[width=0.75\linewidth]{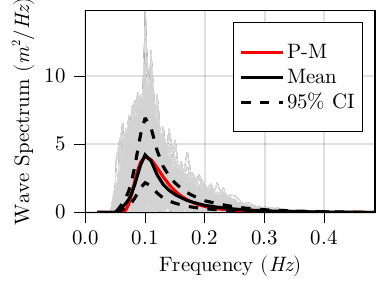}}
        \\
        \subfigure[]{\includegraphics[width=0.75\linewidth]{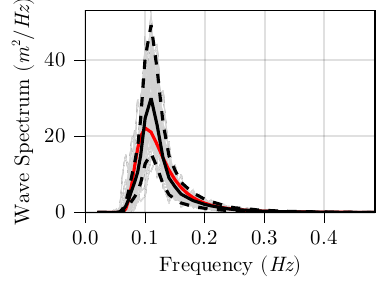}}
        \caption{(a): Illustration of binning method for mean power estimation using wave spectra parameterized by two parameters: significant wave height ($H_s$) and energy period ($T_e$). Bin centers (black dots) are used as the representative spectra. (b): All measured spectra in a bin at the center of the distribution ($H_s = 2.16 m$, $T_e = 8.43 s$). (c): All measured spectra in a bin near the steepness limit of the distribution ($H_s = 5.03 m$, $T_e = 8.43 s$). In both cases the two-parameter Pierson-Moskowitz spectrum (red line, label ``\emph{P-M}'') captures the mean (black line) well, but, specially for the center bin (b) it can be seen that the variability cannot be attributed to the sampling variance (dashed black lines show 95-percentile range) alone.}
        \label{fig:two-parameter}
    \end{figure}

    Wave energy converters (WEC) extract energy from ocean waves---a purely oscillatory resource. 
    This resource is characterized in terms of wave spectra, i.e. spectral density of the variance of the surface elevation.
    In particular, parametric wave spectra are used in engineering analysis to approximate real ocean conditions based on a small number of parameters. 
    Two-parameters spectra, based on significant wave height and energy period (or equivalently peak period), are ubiquitous in practice.  
    The implicit assumption being that all observed spectra at that location with similar values of significant wave height and peak period are different realizations---based on finite records---of the same underlying stationary and ergodic Gaussian process. 
    This assumption is wrong~\cite{SAULNIER2011130,MACKAY201617,merigaud2018power}, as illustrated in Figure~\ref{fig:two-parameter}, and has profound impact on the engineering analysis of WEC performance~\cite{SAULNIER2011130}. 
    
    The common two-parameter spectra, such as JONSWAP~\cite{Hasselmann1973} or Pierson-Moskowitz~\cite{Pierson1964}, were developed for sea states with specific characteristics, e.g., fully developed seas, not for all occurring sea states. 
    This is adequate when considering the performance of engineered structures, such as ships, at specific sea states. 
    In particular, for extreme and steep sea states, spectral shapes do converge based on these two parameters~\cite{young1993review}. 
    However, WECs are different from other ocean applications in that waves are not merely something to tolerate and consider for survivability, but are the main driving design consideration. 
    In particular, WECs are designed to extract energy from the most occurring sea states with extreme sea states playing little or no role in the energy production. 
    The seasonality of the resource also means the performance of the WEC must be evaluated over all possible sea states throughout a year, not just spot checked for a few hand-selected representative or extreme sea states, with the mean annual power production being an adequate metric. 
    
\begin{figure}
        \centering
        \includegraphics[width=0.95\linewidth]{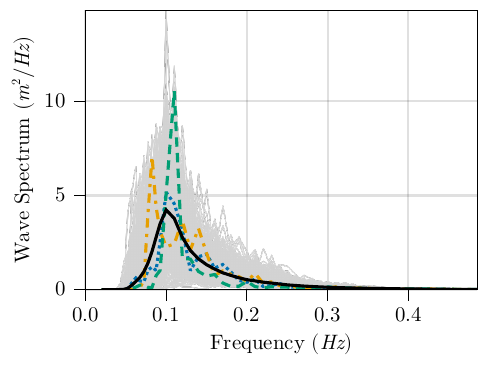}
        \caption{Illustration of the variability in spectral shapes within a single bin ($H_s = 2.16 m$, $T_e = 8.43 s$ in Figure~\ref{fig:two-parameter} (b)). 
        As in Figure~\ref{fig:two-parameter} all spectra are shown in grey and their mean is shown as the thicker black line. 
        Three different spectra are highlighted to illustrate the variability present in these spectra: one spectrum (blue/dots) mostly aligns with the mean, another (orange/dash-dots) contains peaks other than the peak frequency, and the third (green/dashes) has a narrower energy bandwidth concentrated near the peak frequency.}
        \label{fig:vary_spectra}
    \end{figure}

    The typical approach to evaluating mean annual power is the \emph{method of bins} (e.g., ~\cite{iec-100}) which consists of (i) obtaining and parameterizing wave spectra records, (ii) binning the data in parameter space, (iii) evaluating the power performance of the device at one representative sea state per bin, and (iv) performing a weighted sum based on the likelihood of each bin. 
    The records in step (i) must contain whole years to avoid seasonality bias and be long enough for statistical properties to converge. 
    Figure~\ref{fig:two-parameter} depicts this process, showing the parameter-space bins and the bin centers as representative sea states. 
    It is clear from Figure~\ref{fig:two-parameter} that the assumption that these two parameters describe an underlying stationary and ergodic random process for all sea states in a given bin is implicit in this approach for estimating mean annual power. 
    Figure~\ref{fig:vary_spectra} illustrates the range of spectral shapes present in a single bin, something that cannot be captured by traditional two-parameter spectra.

    Several methods have been developed to characterize sea states other than fully-developed or fetch limited sea states or to describe all possible sea states.  
    These include multi-modal parametric sea states like Ochi-Hubble spectrum~\cite{ochi1976six}, for which the spectral shape varies and depends on the significant wave height and energy period. 
    Although in these models the shape is a function of the parameters, these are still two-parameter spectra and cannot not provide an adequate representation of all sea states~\cite{MACKAY201617}. 
    Another approach studied extensively~\cite{tracy2007wind} is partitioning the sea state into wind seas and swell, and fitting different spectra to each. 
    Yet another approach is to create a single parametric spectra that can capture all sea states relatively well. 
    This approach was undertaken by Mackay~\cite{MACKAY201617} who developed a 4-parameter spectra shown to perform well over a large number of sea states. 
    Mackay's spectra is based on the averaging of two JONSWAP spectra and the two shape parameters are obtained from analysing the distribution of energy about the energy period. The expression of the two parameters is not closed-form and requires the use of table look-up and interpolation based on buoy data from 8 different locations.  
    
    While Mackay's spectra is shown to capture sea state variability well, the effect of using such a spectra for mean annual power of WECs has not been studied. 
    Using sensitivity analysis, Saulnier et al.~\cite{SAULNIER2011130} showed the insufficiency of significant wave height and energy period in modeling the performance of WECs, and demonstrated that in some cases complementing these two parameters with a bandwidth parameter is sufficient for describing the sea state for the purpose of WEC power production.
    However, no practical method exists to incorporate this, e.g., in the form of a 3-parameter spectra, into the mean annual power analysis. 
    Since 2012 the IEC standards explicitly call out the possibility of using additional parameters, such as wave direction and spectral bandwidth~\cite{iec-100}, for the bins to reduce the variability of power performance within bins. 
    However, as far as the authors are aware, there are no such studies in the open literature. 
    
    In this study we evaluate the effect of using the 4-parameter Mackay spectra in an mean annual power analysis and compare it to using the 2-parameter Pierson-Moskowitz~\cite{Pierson1964} spectra. 
    We also develop and demonstrate a machine learning--based 4-parameter spectra using an autoencoder architecture as a way to develop site-specific parametric spectra from measured data.
    Autoencoders~\cite{Rumelhart_1988, Goodfellow-et-al-2016} are deep neural network architectures capable of learning efficient encoding of data and can be used to create reduced order representations. 
    We augment the traditional autoencoder architecture to enforce known physical properties, a process known as physics-informed machine learning, to fix the significant wave height and energy period, and allow the autoencoder to focus solely on shape parameters.  
    We compare the predicted annual mean power for two case studies.
    The 4-parameter spectra are shown to be more consistent for modeling WEC mean annual power, with the site-specific autoencoder-based spectra resulting in the most consistently accurate results.

\section{Methodology}
    \subsection{WEC Model}
\begin{figure}
            \centering
            \def\svgwidth{\linewidth}
            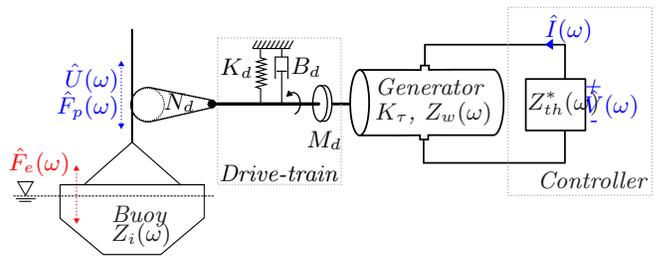
            \caption{Linear wave-to-wire model for the WaveBot device. The frequency dependence is expressed in terms of the radial frequency $\omega=2\pi f$. The optimal controller is used, where the load impedance is set to the complex conjugate of the Thévenin impedance. The model is described in Appendix~\ref{App:WaveBot}.}
            \label{fig:wavebot}
        \end{figure}

        We use the well-studied WaveBot~\cite{Coe2016a} WEC, which is a small, axisymmetric device with a waterplane radius of $0.88\,m$ and a total depth of $0.53\,m$. 
        We consider the device motions and power extraction in one degree of freedom: heave. 
        The frequency--domain linear wave-to-wire model~\cite{michelen2023control}, illustrated in Figure~\ref{fig:wavebot}, models the average electric power production of the device at a given sea state. 
        The model assumes optimal control strategy based on impedance matching using the Th\`{e}venin equivalent circuit.
        The model is presented in detail in~\cite{michelen2023control} and summarized in Appendix~\ref{App:WaveBot}. 
    
    \subsection{Wave Resource Data}
        A real sea state is characterized by its wave spectrum $S(f, \theta)$: the spectral density of the variance of sea surface elevation with respect to frequency $f$ and wave direction $\theta$. 
        Since the WaveBot is axisymmetric we can consider the omnidirectional spectrum: 
        \begin{equation}
            S(f) = \int_0^{2\pi}S(f, \theta)d\theta. \label{eq:omni} 
        \end{equation}
        As a deployment location we use the PacWave~\cite{pacwaveReport} site, and obtain buoy measurements from NOAA's NDBC~\cite{ndbc} buoy 46050. 
        These are 30-minute omnidirectional wave spectra estimates reported every hour.
        The buoy spectra are calculated at $N=47$ unevenly spaced frequencies, and capture both wind seas and swells. 
        Because sea states have seasonal variations, the total period of record must be an integer number of years, and must be long enough (number of years) for statistical properties to converge.
        Years with missing periods (e.g. due to damage or other downtime of the buoy) were discarded as to not add bias due to seasonality effects. 
        In total we kept 11 full years between 2009--2021.
        We further remove spectra with unrealistic steepness $s_{e} > 0.1$ and excessive low frequency energy, $S(f) > 0.01m^2/Hz$ for $f <= 0.0325 Hz$, based on Mackay \cite{MACKAY201617}.
        The significant steepness, $s_e$, for deep water is expressed as
        \begin{equation}
            s_e = \frac{2\pi H_s}{gT_{e}^2}
        \end{equation}
        where $g$ is the gravitational acceleration. 
        In total, considering missing and removed data, we retain 93,426 out of 96,432 spectra for the 11 years retained, which corresponds to 96.88\% of the data. 
        Table~\ref{tab:waves} gives a summary of the data used. 

        \begin{table}
            \centering
            \caption{Usable wave spectra from NDBC Buoy 46050 per year and mean annual power estimates (based on all retained spectra) used as ground truth.}
            \begin{tabular}{rccc}
                \toprule
                \textbf{Year} & \textbf{No. Spectra} & \textbf{\% Spectra} & \textbf{Avg. Power}\\
                \midrule
                2009 & 8,396 & 95.84\% & 1,337 W \\
                2010 & 8,375 & 95.61\% & 1,721 W \\
                2012 & 8,388 & 95.49\% & 1,506 W \\
                2013 & 8,418 & 96.10\% & 1,076 W \\
                2014 & 8,561 & 97.73\% & 1,313 W \\
                2016 & 8,597 & 97.87\% & 1,515 W \\
                2017 & 8,622 & 98.42\% & 1,334 W \\
                2018 & 8,592 & 98.08\% & 1,246 W \\
                2019 & 8,461 & 96.59\% & 1,172 W \\
                2020 & 8,489 & 96.64\% & 1,355 W \\
                2021 & 8,527 & 97.34\% & 1,529 W \\
                \bottomrule
                \textbf{} & \textbf{93,426} & \textbf{96.88\%} & \textbf{1,373 W} \\
                \bottomrule
            \end{tabular}
            \label{tab:waves}
        \end{table}
        
        \begin{figure}
            \centering
            \includegraphics[width=0.5\linewidth]{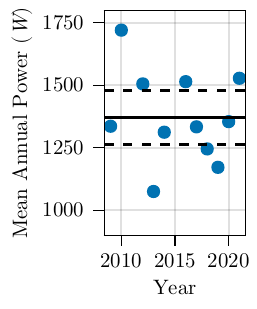}%
            \caption{Mean annual power calculated directly from all measured spectra in Table~\ref{tab:waves} and used as ground truth. Dashed lines show 95\textsuperscript{th} percent confidence interval for the mean (solid line).}
              \label{fig:annual_power}
        \end{figure}

        The spectra from the buoys are subject to sampling error due to the finite record length and the measurement and processing methodology~\cite{ndbc}. 
        The Sampling error can be described using a $\chi^2$-distribution, with $d=24$ degrees of freedom for this specific type of NDBC buoy~\cite{ndbc,mackay2011modelling}. 
        For a specific underlying spectrum $\overline{S}(f)$ the distribution of samples $S_i(f)$ is given as
        \begin{equation}
            S_i(f) \sim \frac{1}{d}\overline{S}(f)\chi^2(d). \label{eq:xi}
        \end{equation}
        The distribution  in \eqref{eq:xi} is used to plot the 95\% confidence interval, e.g., in Figure~\ref{fig:two-parameter}, for a subset of the spectra with $\overline{S}$ being the mean of the subset.
        This serves as a visual indication of whether the subset can be considered different samples of the same underlying stationary and ergodic Gaussian process.
        
    \subsection{Mean Annual Power}
        The mean annual power for each year of data can be calculated directly by running the WaveBot model for each spectra.
        When possible, this is the most accurate approach, and we use it here as a synthetic \emph{ground truth}. 
        We average the yearly estimates to obtain a mean annual power of $1,373\,W$ as shown in Table~\ref{tab:waves} and Figure~\ref{fig:annual_power}.
        In practice however, this is prohibitively expensive when using more accurate non-linear time-domain models or wave tank testing. 
        Instead, the less time-consuming \emph{method of bins} is generally used. 
        The methods of bins is as follows: 

        \begin{algorithm}
            \item Represent all measured wave spectra in terms of a small set of parameters, typically significant wave height ($H_s)$ and energy period ($T_e$).
            \item Bin the parameter space.
            \item Create a representative spectra for each bin, using the parameters at the bin centers and a parametric spectrum.
            \item Estimate the WEC's average power at the representative sea states, e.g. using a numerical or physical model.
            \item Sum the powers weighted by the proportion of data within each bin, i.e. the relative frequency of each bin.
        \end{algorithm}
        
        The parameter-space binning process is illustrated in Figure~\ref{fig:two-parameter}. 
        The two parameters---significant wave height and energy period---are defined in terms of spectral moments $m_n$ as 
        \begin{align}
            H_s &= 4\sqrt{m_0} \\ 
            T_e &= \frac{m_{-1}}{m_0} \\
            m_n = \int_0^\infty f^n & S(f) df \approx \sum_{i=1}^N f_i^n S(f_i)
        \end{align}
        and can be easily calculated for each measured spectrum. 

        A parametric wave spectrum is needed to go from parameter space to wave spectrum for the representative sea states, i.e., bin centers. 
        For the PacWave site the Pierson-Moskowitz spectrum~\cite{Pierson1964} is an appropriate two-parameter spectrum~\cite{pacwaveReport}. 
        The Pierson-Moskowitz spectrum, for a given value of $H_s$ and $T_e$ is given as
        \begin{equation}
            S(f;\,H_s, T_e) = \frac{5}{16}H_s^2f^{-1}\left(f_p/f\right)^4 e^{\left[-\frac{5}{4}\left(f_p/f\right)^4\right]}
        \end{equation}
        where the peak frequency is $f_p = 0.858/T_e$.

        We note that the method used here is unrealistic for several reasons. 
        First, the WaveBot is not designed for power production for real ocean waves. 
        Additionally, the linear model is not accurate for large waves or large WEC motions and therefore should not be used for all possible waves as we do here. 
        Finally, a real WEC would likely not operate at all possible sea states, e.g., extreme sea states. 
        However, because of its low cost, the linear model allows us to create a synthetic ground truth to compare to and evaluate the effectiveness of the method of bins.
        
    \subsection{Spectral Shape and Scaling}
        When considering additional parameters, it is convenient to \emph{normalize} out the significant wave height and the energy period. 
        The significant wave height is a measure of the \emph{magnitude} of elevation variance or total energy while the energy period is a measure of the \emph{location}. 
        The resulting non-dimensional spectrum is considered the \emph{spectral shape} and additional parameters can be developed to represent this shape. 
        This is the approach taken by Mackay~\cite{MACKAY201617} and for the autoencoder-based spectrum developed here. 
        The normalized spectrum $\Tilde{S}$ is given as 
        \begin{align}
            \Tilde{S}(\Tilde{f}) &= S(f) \frac{1}{H_s^2T_e} \label{eq:nondim_s}\\
            \Tilde{f} &= f T_e\label{eq:nondim_f}.
        \end{align}
        In general we can re-scale a spectrum to have the same spectral shape but different $H_s$ and $T_e$ as 
        \begin{align}
            S(f) &= S_o(f_o)\frac{H_s^2T_e}{H_{s,o}^2T_{e,o}} \label{eq:rescale_s}\\
            f &= f_o\frac{T_e}{T_{e,o}}\label{eq:rescale_f}.
        \end{align}
        with the subscript $o$ indicating the original values. 

\begin{figure}
            \centering
            \includegraphics[width=0.95\linewidth]{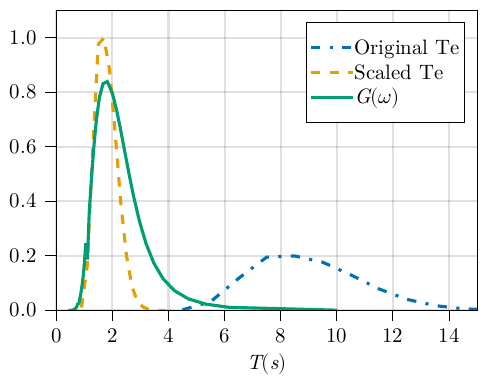}
            \caption{Device performance (power gain $G$) compared to energy distribution of the resource (energy period $T_e$) for both the measured sea states and the scaled sea states (energy period scaled by a factor of $5$). 
            The scaled sea states provide a second case study that considers a better match between the device performance and the resource.}
            \label{fig:Ga}
        \end{figure}

        We use these relationships to consider scaled-down sea states that represent a more realistic match between the WaveBot device and the resource. 
        The WaveBot was not designed for deployment in real seas and is badly tuned for extracting energy from the PacWave site. 
        This can be seen in Figure~\ref{fig:Ga} where the device's power gain $G$ is compared to the distribution of energy period. 
        The power gain, or \emph{wave-to-wire efficiency}, is the ratio of power produced by the WEC to the power entering the WEC, and is described in Appendix~\ref{App:WaveBot}. 
        It can be seen that the WaveBot is most efficient at frequencies not found in the deployment site. 
        For this reason we consider the case of a scaled sea states, in addition to the real sea states.

        We scale the sea states to have a new energy period of $T_e = \alpha T_{e,o}$ with $\alpha=1/5$. 
        This means that the wavelengths are scaled by a factor of $\alpha^2=1/25$ based on the deep water dispersion relation. 
        For geometric similitude, we scale the significant wave height by this same factor of $\alpha^2$.
        The total energy in the waves is therefore scaled by a factor of $\alpha^4=1/625$. 
        The distribution of $Te$ for the scaled sea states is also shown in Figure~\ref{fig:Ga} and it can be seen that the device performance and the resource are better matched. 

    \subsection{Mackay Spectrum}
        The Mackay spectrum~\cite{MACKAY201617} describes the spectral shape based on two additional parameters. 
        The idea is to describe the distribution of energy about the energy period. 
        To obtain the two parameters from a spectrum, first the spectrum is partitioned about $f_e = 1/T_e$ and the significant wave height and energy period of each partition is calculated as
        \begin{align}
            & H_{s,A} = 4\sqrt{m_{0,A}} & H_{s,B} = 4\sqrt{m_{0,B}}\\
            & T_{e,A} = \frac{m_{-1,A}}{m_{0,A}} & T_{e,B} = \frac{m_{-1,B}}{m_{0,B}} \\
            & m_{n,A} = \int_0^{f_e}S(f)f^n df & m_{n,B} = \int_{f_e}^\infty S(f)f^n df
        \end{align}
        The significant wave height and energy period of the two partitions and those of the overall spectra are related so that the number of parameters can be reduced from 4 to 2. 
        The two parameters used by Mackay are the normalized significant wave height of the first partition squared, $H_{An}^2$ which is proportional to the total energy in the first partition, and the \emph{bimodality parameter}, $dT_n$ which is the difference of the normalized energy periods: 
        \begin{align}
            H_{An}^2 &= \left(\frac{H_{s,A}}{H_s}\right)^2 \\
            dT_n &= \frac{T_{e,A}}{T_e} - \frac{T_{e,B}}{T_e}.
        \end{align}

        Mackay~\cite{MACKAY201617} formulates a 4-parameter spectrum as a sum of two separate JONSWAP spectra. 
        \begin{equation}
            S(f;\,H_s, T_e, H_{An}^2, dT_n) = 
            \sum_{i=1}^2 S_{J}(f;\,H_{s,i}, T_{e,i}, \gamma_i)
        \end{equation}
        where $S_J$ is the standard JONSWAP spectrum. 
        The relationship between the two shape parameters and the 6 parameters for the two JONSWAP spectra are determined empirically by interpolating from 6 different look-up tables provided by Mackay. 
        The empirical data is based on NDBC data from 8 different deep water buoys: 5 in the north pacific, one in the Gulf of Mexico, one in the Caribbean Sea, and one in the North Atlantic. 
        Buoy 46015 is in the Pacific Northwest coast of the United States, close to the PacWave site we consider in this study. 
        The spectra were partitioned and fitted, a computationally expensive process, to create the lookup tables. 
        The JONSWAP parameters are given as  
        \begin{align}
            & H_{s,1} = H_s \cdot h_{1,n}(H_{An}^2, dT_n) & H_{s,2} = H_s \cdot h_{2,n}(H_{An}^2, dT_n) \\
            & T_{e,1} = T_e \cdot t_{1,n}(H_{An}^2, dT_n) & T_{e,2} = T_e \cdot t_{2,n}(H_{An}^2, dT_n) \\
            & \gamma_1 = \gamma_1(H_{An}^2, dT_n) & \gamma_2 = \gamma_2(H_{An}^2, dT_n)
        \end{align}
        where $h_{1,n}$, $h_{2,n}$, $t_{1,n}$, $t_{2,n}$, $\gamma_1$, and $\gamma_2$ are the six interpolation functions.

    \subsection{Autoencoder Spectrum}
        We use an autoencoder, augmented with a physics-informed layer, to obtain the two shape parameters ($\theta_1$, $\theta_2$) in our 4-parameter parameterization: 
        \begin{equation}
            S(f;\,H_s, T_e, \theta_1, \theta_2) . \label{eq:autoenc}
        \end{equation}
        The decoder represents the scaled spectra (shape) $\tilde{S}$ and \eqref{eq:autoenc} can be expressed as
        \begin{equation}
            S(f;\,H_s, T_e, \theta_1, \theta_2) = H_s^2T_e\tilde{S}(\tilde{f};\,\theta_1, \theta_2),
        \end{equation}
        where $\tilde{f}=fT_e$.
        
        \subsubsection{Autoencoder Architecture}
        Autoencoders are a type of deep neural network architecture capable of learning efficient encodings of data. 
        They often have a symmetrical architecture, where the first half of the neural network---the encoder---consists of a series of layers that create a lower dimensional output while, conversely, the other half of the model---the decoder---transforms the lower dimensional encoding back to its original dimension. 
        In this framework, the training data inputs and outputs are the same, i.e. the optimizer attempts to set the network's parameters so that the outputs are as close as possible to the inputs, while propagating through a network architecture that forces reduction to a low dimensional space. 
        This results in the desired encoding that preserves as much information as possible. 
        This is fundamentally different than classical neural network architectures used to learn a functional mapping between inputs and outputs, where the training data consists of input-output pairs and creation of the training data outputs often requires intensive human labor.

        \begin{figure}
            \centering
            \includegraphics[width=\linewidth]{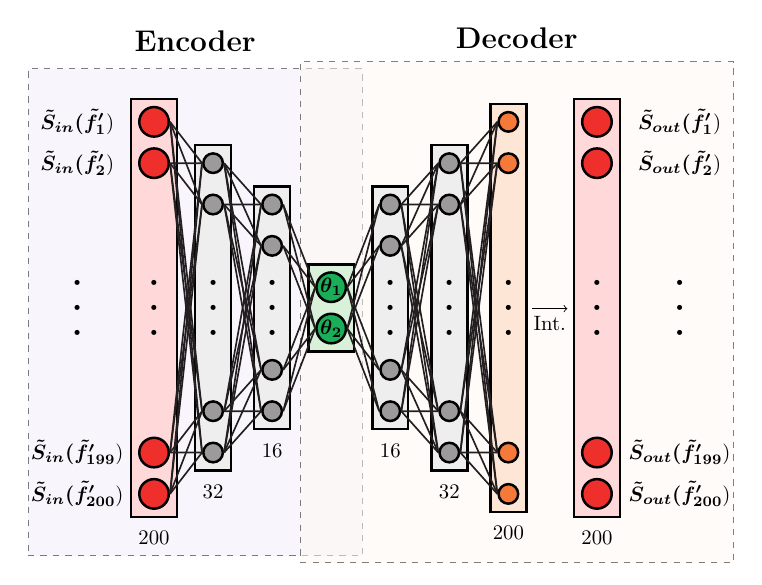}
            \caption{Autoencoder architecture. 
            A LeakyReLU activation function is used for most layers except when mentioned otherwise. 
            The encoder takes a normalized spectrum $\tilde{S}$ (i.e., $H_s=1$, $T_e=1$) at $N = 200$ predefined normalized frequencies $\tilde{f}'$ evenly spaced between [0.01, 7.5] as inputs, and outputs the two shape parameters, using a Sigmoid activation function for the last encoding layer (green). 
            The decoder takes the two shape parameters as inputs and returns the normalized spectrum, using a modified Softmax activation function to enforce $H_s=1$ and $T_e=1$ (orange). 
            The use of the Softmax layer changes the normalized frequencies, and an interpolation is required to return to the same $N = 200$ frequencies as in the input of the encoder (red). 
            The trainable parameters consists of all black lines (weights) and non-red inner nodes (biases).  
            The training optimizes these parameters such that the output normalized spectrum is as close as possible to the input ($\tilde{S}_{out}\approx\tilde{S}_{in}$).
            }
            \label{fig:autoencoder}
        \end{figure}
        
        The autoencoder architecture we use here is shown in Figure~\ref{fig:autoencoder}. 
        The majority of layers use a Leaky ReLU activation function, except when specified otherwise.
        The input to the encoder consists of a normalized spectrum $\tilde{S}_{in}$ at $200$ predefined normalized frequencies $\tilde{f}'$ evenly spaced between [0.01-7.5]. 
        A larger number of frequencies ($N=200$) is used compared to those from the measured data ($N=47$) in order to reduce the errors from discretization during different interpolations steps discussed below. 
        The outputs of the encoder are the two shape parameters ($\theta_1$, $\theta_2$). 
        We use a Sigmoid function for the last encoding layer to provide bounds to the two shape parameters, limiting both to the range $[0,1]$.
        While not required, without these bounds every retraining of the network result in arbitrary parameter ranges that can be orders of magnitude different. 
        Since the magnitude of these parameters has no physically meaningful interpretation it makes sense to have them take values within this bounded range. 
        
        The decoder inputs consists of the two shape parameters and its outputs are the normalized spectrum $\tilde{S}_{out}$ at the same $200$ normalized frequencies as for $\tilde{S}_{in}$.
        After the last trainable decoder layer (orange) we use an activation function that acts similar to a Softmax function and enforces the hard constraints $H_s=1$ and $T_e=1$, described in the next subsection. 
        Normalization using equation~\eqref{eq:nondim_f} results in a new set of frequencies, and as a last step we interpolate back to the common set of normalized frequencies $\tilde{f}'$. 

        For training the autoencoder we use stochastic gradient descent with batch size of 512 spectra, and an Adam optimizer~\cite{ADAM} at a learning rate of $\eta=1*10^{-4}$. 
        Because the model is attempting to optimize solely spectral shape, we minimize the difference between $\tilde{S}_{in}$ and $\tilde{S}_{out}$ using the root mean square error (RMSE) loss~\eqref{eq:rmse}. 
        Following Figure~\ref{fig:autoencoder}, we are trying to minimize the difference of the data at the red layers. 
        The loss function for a single spectra is given as
        \begin{equation}
            \mathcal{L} = \sqrt{\frac{1}{N}\sum_{i=1}^{N}(\tilde{S}_{out,i} - \tilde{S}_{in,i})^2}
            \label{eq:rmse}
        \end{equation}

        \subsubsection{Physics-Informed Machine Learning}
        Physics-informed machine learning (PIML)~\cite{karniadakis2021physics} refers to methods to embed known physics into machine learning processes. 
        Many of the advantages of PIML are derived from the fact that the learning algorithm needs to learn less, since part of the relationship is embedded and it does not need to ``reinvent the wheel". 
        In practice this means being able to train with less, even sparse, data, accelerate training, and enhance generalization of the learned function. 
        In physics-based problems we often don't have ``Big Data" but conversely often know physical symmetries or conservation laws that don't need to be relearned. 
        PIML can be implemented as either soft or hard constraints. 
        Examples of soft constraints include \emph{observational bias}, i.e., artificially expanding the training data to bias the training with known relationships, and \emph{learning bias} where the residuals of known physical relationships are included as additional terms to the objective function of the optimizer. 
        These soft constraints do not strictly enforce the known physics, but bias the training process such that these relationships generally hold. 
        In contrast, hard constraints, or \emph{inductive bias}, automatically and strictly enforce the known physical relations through specialized design of the network architecture. 
        
        The goal of our autoencoder is to parameterize spectral shapes. 
        As such we want the encoding and decoding process to preserve both the total energy and location, and all error in the encoding/decoding process to be attributable to differences in spectral shape. 
        We therefore chose to enforce preservation of $H_s$ and $T_e$ through hard constraints, based on the architecture design of the neural network.
        These constraints are enforced using~\eqref{eq:nondim_s} and~\eqref{eq:nondim_f} in the initial output layer (shown in orange in Figure~\ref{fig:autoencoder}) resulting in an activation function analogous to a Softmax function. 
        The use of~\eqref{eq:nondim_f} results in a modified set of normalized frequencies $\tilde{f}$, and as a last step we interpolate back into the same set of $200$ frequencies used in the input to the encoder. 

        \subsubsection{Parametric Spectra}
        The entire process of parameterizing a real sea state, or conversely creating a sea state from a set of parameters, encompasses a few steps beyond the use of the trained autoencoder in Figure~\ref{fig:autoencoder}. 
        As a first step before calling the trained encoder, a given discrete spectrum must be normalized and interpolated onto the predefined set of input frequencies. 
        An evenly spaced grid of 200 non-dimensional frequencies were chosen in the range [0.01, 7.5] that covered the majority of the range seen in all the training $\tilde{f}'$.
        This was a compromise between balancing training time and the risk of losing information through the interpolation process. 
        The risk of losing information is due to the values of $\tilde{f}$ having different ranges for different sea states. 
        Any given spectrum, once normalized, covers a subset of the 200 normalized frequencies, and the value of the normalized spectrum at either end can be simply extrapolated to zero.  

        Similarly, when using the decoder, the output spectrum must be redimensionalized based on $H_s$ and $T_e$ using~\eqref{eq:nondim_s} and~\eqref{eq:nondim_f}. 
        The output discrete spectra are all at the same non-dimensional frequencies, but at different dimensional frequencies. 
        After re-dimensionalizing, we can optionally interpolate into a common set of dimensional frequencies. 
        For our power calculation studies we interpolate back into the same set of $47$ frequencies used by the NDBC buoy.
        
        We wrap all these steps into a single parameterization function (encoding process) that takes a real sea state from the NDBC buoy and gives the 4 parameters ($H_s$, $T_e$, $\theta_1$, $\theta_2$), as follows:

        \begin{algorithm}[\alph*)]
            \item \textbf{INPUT:} Start with a discrete spectrum $S(f)$ at N frequencies.
            \item Calculate $H_s$ and $T_e$ and normalize the spectrum to $\tilde{S}(\tilde{f})$ using \eqref{eq:nondim_s} and \eqref{eq:nondim_f}.
            \item Interpolate these to the pre-defined set of common non-dimensional frequencies $\tilde{f}'$. 
            \item Pass the data $\tilde{S}(\tilde{f}')$ into the encoder to obtain $\theta_1$ and $\theta_2$.
            \item \textbf{OUTPUT:} four parameters representing the spectrum: ($H_s$, $T_e$, $\theta_1$, $\theta_2$).
        \end{algorithm}

        The converse process, creating a spectra from given parameters, is similarly wrapped into a single parametric spectra function that internally takes the following steps:
        \begin{algorithm}[\alph*)]
            \item \textbf{INPUT:} The four parameters that represent the spectrum: ($H_s$, $T_e$, $\theta_1$, $\theta_2$)
            \item Pass the shape parameters ($\theta_1$, $\theta_2$) through the decoder to obtain a normalized spectra at the pre-defined set of common non-dimensional frequencies $\tilde{f}'$.
            \item Using \eqref{eq:rescale_s} and \eqref{eq:rescale_f} re-scale the spectra using $H_s$ and $T_e$. 
            \item Optional: Interpolate the spectra to a prefered set of dimensional frequencies.
            \item \textbf{OUTPUT:} Discrete spectrum $S(f)$.
        \end{algorithm}

\section{Results}
    We present the results of calculating mean annual power for two case-studies: the WaveBot operating at the measured sea state at the study site, and the same WaveBot operating at the scaled sea state discussed earlier, where the energy period of each sea-state is scaled by $\alpha=1/5$. 
    The scaled sea-state case provides a more realistic match between WEC design and resource. 
    
    We use the method of bins to find the corresponding parameters at evenly spaced points and generate representative spectra. 
    For the mean annual power calculation, we explore the convergence of the method of bins as the number of bins increases. 
    We use a simple discretization with the same number of bins, $N$, for each dimension (parameter). 
    For a given discretization $N$ a two-parameter spectra would have $N^2$ bins and a 4-parameter spectra would have $N^4$ bins.
    We also plot the limit, which is the mean annual power using the parametric spectra for every sea state in the data ($93,426$ spectra). 
    We start with relatively few number of bins in order to show the convergence.

    \subsection{Autoencoder Training}
\begin{figure}[ht]
        \centering
        \subfigure[]{\includegraphics[width=0.95\linewidth]{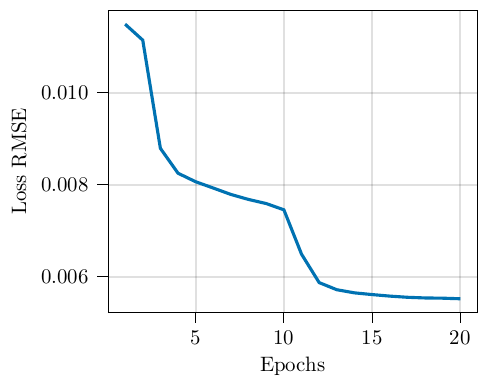} \label{fig:training_loss}}
        \subfigure[]{\includegraphics[width=0.95\linewidth]{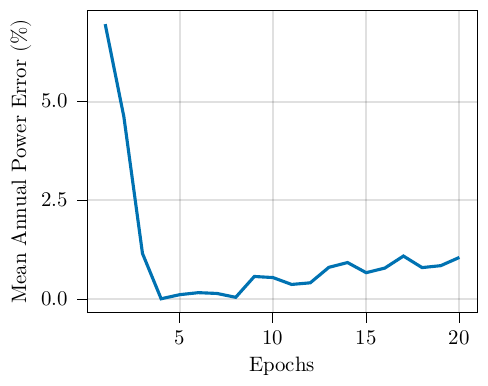} \label{fig:training_pct}}
        \caption{Autoencoder training process. (a) Loss function, root mean squared error, during training. (b) Percent error in power prediction for the scaled waves case for each training epoch. The error in power prediction is not part of the training, and shows the onset of overfitting.}
        \label{fig:loss_graph}
    \end{figure}

    Using a computer with an AMD Ryzen 7 5700X 8-Core Processor, with 32 GiB of available RAM, each epoch took approximately 110-115 minutes.
    Totalling around 40 hours to train completely for 20 epochs.
    Figure~\ref{fig:loss_graph} illustrates the training process. 
    Figure~\ref{fig:training_loss} shows the training loss equation~\eqref{eq:rmse}, which is the value the optimizer is attempting to minimize.  
    Figure~\ref{fig:training_pct} shows the percent error in predicted annual power for the scaled waves case. 
    The power is not used in training but shows the onset of overfitting. 
    In this case, at a certain point further decrease in training error can result in larger errors in our quantity of interest (power). 

    \subsection{Autoencoder Spectra}
\begin{figure}[ht]
        \centering
        \includegraphics[width=0.95\linewidth]{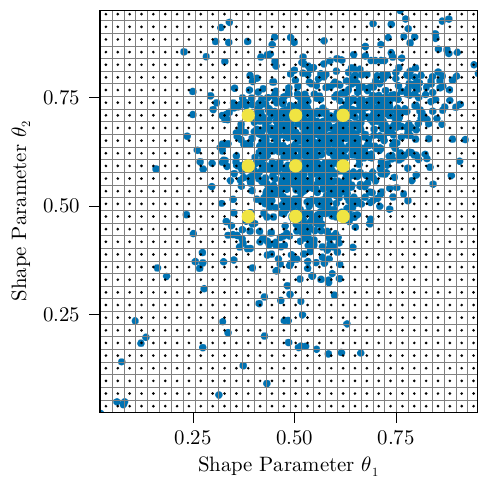}
        \caption{Shape parameter bins for $H_s=2.16 m$ and $T_e = 8.43 s$ and 9 evenly spaced selected bins shown in yellow.}
        \label{fig:latent_var_grid}
    \end{figure}

\begin{figure*}[ht]
        \centering
        \includegraphics[width=0.95\linewidth]{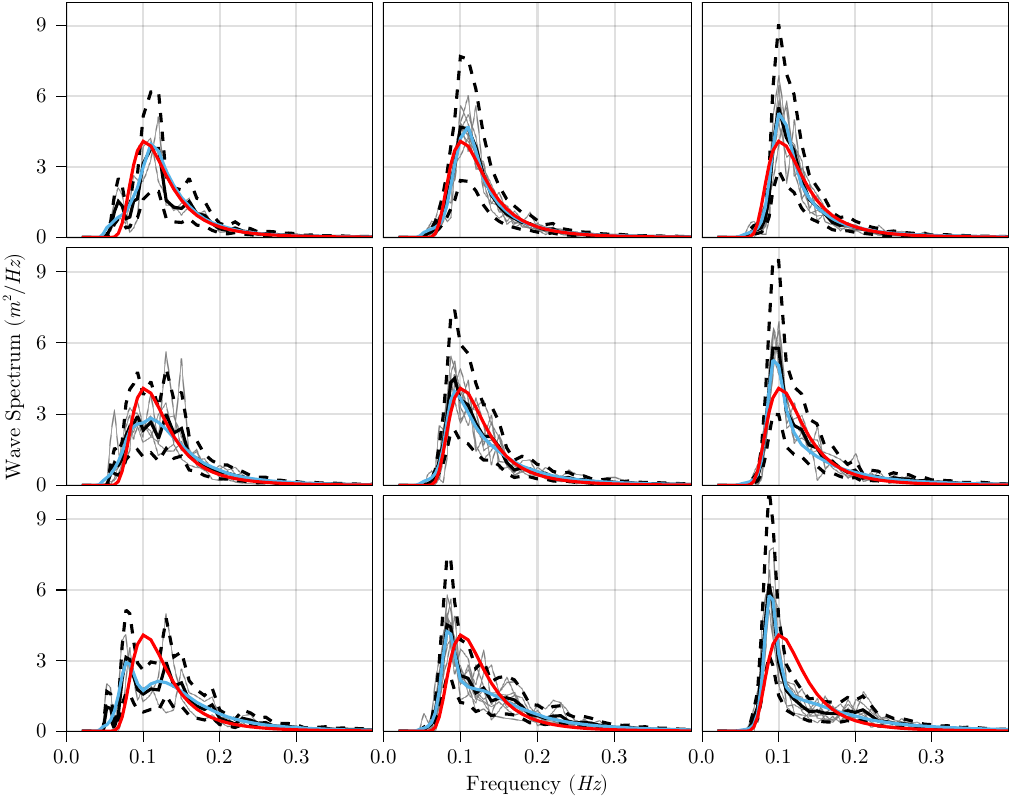}
        \caption{Variability within different 4-parameter bins, for the 9 bins in Figure~\ref{fig:latent_var_grid}. Samples are shown in grey, the sample mean in solid black lines, and the 95\% confidence interval in dashed black lines. The autoencoder and Pierson-Moskowitz spectra are shown in light blue and red solid lines, respectively. }
        \label{fig:latent_grid_ex}
    \end{figure*}

    We now look at whether the trained autoencoder spectra is an adequate representation of the sea state, i.e., whether the variability of all samples within a bin can be attributed to the sampling variance. 
    In this case a given $H_s$--$T_e$ bin is subdivided into multiple $\theta_1$--$\theta_2$ bins. 
    We look at the same bin as in Figure~\ref{fig:vary_spectra} ($H_s = 2.16 m$, $T_e = 8.43 s$) and choose 9 sub-bins ($\theta_1$, $\theta_2$) as illustrated in Figure~\ref{fig:latent_var_grid}. 
    Figure~\ref{fig:latent_grid_ex} shows the large variability of spectral shapes even for similar values of $H_s$ and $T_e$, including some multimodal and skewed spectral shapes. 
    Qualitatively, we can say that the variability of the samples within each bin can be attributed to sampling variance, with most samples falling within the $95\%$ confidence interval.
    This suggests the use of these 4-parameters is adequate to capture spectra variability. 
    The autoencoder spectra captures the mean of the samples well for each bin, whereas a single Pierson-Moskowitz spectrum does not. 
    This shows that the use of 4-parameters and the autoencoder spectra is adequate for representing the variability of sea states. 
    
    \subsection{Mean Annual Power---Measured Waves}
\begin{figure}[ht]
        \centering
        \includegraphics[width=0.95\linewidth]{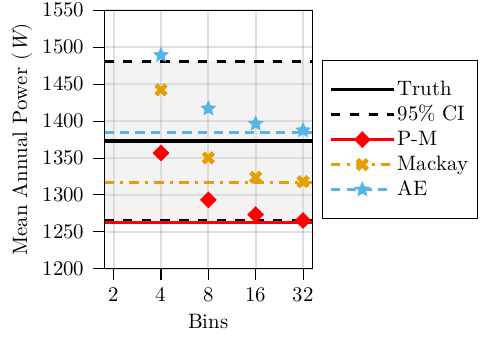}
        \caption{Mean annual power estimates for the \emph{measured} sea state case. The figure shows the convergence of the method of bins for increasing discretization. The limit, treating each sample as its own bin, is shown as solid lines.}
        \label{fig:res_unscaled}
    \end{figure}

    The results of mean annual power estimate using the different parametric spectra are shown in Figure~\ref{fig:res_unscaled} and Table~\ref{tab:unscaled_results}. 
    The autoencoder spectra performs the most accurately with respect to the ground truth, with an error of less than $1\%$. 
    Table \ref{tab:unscaled_results} shows the values of the limit for each method as well as the error when compared to the truth, expressed as a percentage.
    The baseline, Pierson-Moskowitz underestimates the annual power by about 8\%. 
    The Mackay spectra performs somewhat in-between these two with an error of $4\%$ and well within the $95\%$ confidence interval. 
    
    \begin{table}[ht]
        \centering
        \caption{Mean annual power and error for the \emph{measured} sea state using all available spectra (limit of method of bins).}
        \begin{tabular}{|c|c|c|}
        \hline
            Method & Value & Percent Error\\
            \hline
            Truth & 1,373.07 & - \\
            Pierson-Moskowitz & 1,262.24 & -8.07\%\\
            Mackay & 1,317.05 & 4.08\%\\
            Autoencoder & 1,384.24 & 0.81\%\\
            \hline
        \end{tabular}
        \label{tab:unscaled_results}
    \end{table}

    \subsection{Mean Annual Power---Scaled Waves}
\begin{figure}[ht]
        \centering
        \includegraphics[width=0.95\linewidth]{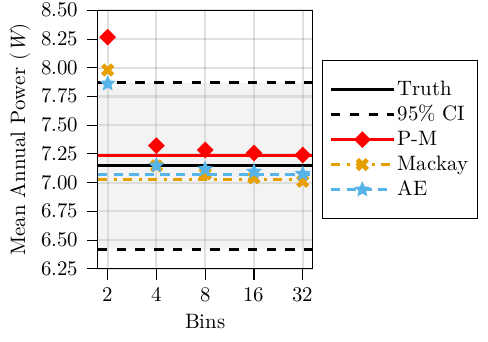}
        \caption{Mean annual power estimates for the \emph{scaled} sea state case. The figure shows the convergence of the method of bins for increasing discretization. The limit, treating each sample as its own bin, is shown as solid lines.}
        \label{fig:res_scaled}
    \end{figure}

    We repeat the same experiment with the scaled sea state that represent a better alignment between the resource and the WaveBot's performance.
    The results are shown in Figure~\ref{fig:res_scaled} and Table~\ref{tab:scaled_results}.
    All three spectra result in very low errors, within $2\%$. 
    It is surprising that the Pierson-Moskowitz performs so well considering how it fails to capture variations in spectral shape within a bin. 
    The following section investigates this further.    
    
    \begin{table}[ht]
        \centering
        \caption{Mean annual power and error for the \emph{scaled} sea state using all available spectra (limit of method of bins).}
        \begin{tabular}{|c|c|c|}
        \hline
            Method & Value & Percent Error\\
            \hline
            Truth & 7.15 & -\\
            Pierson-Moskowitz & 7.23 & 1.23\%\\
            Mackay & 7.03 & -1.68\%\\
            Autoencoder & 7.07 & -1.05\%\\
            \hline
        \end{tabular}
        \label{tab:scaled_results}
    \end{table}
    
    \subsection{Error Distribution}
\begin{figure}[!ht]
        \centering
        \subfigure[]{\includegraphics[width=0.95\linewidth]{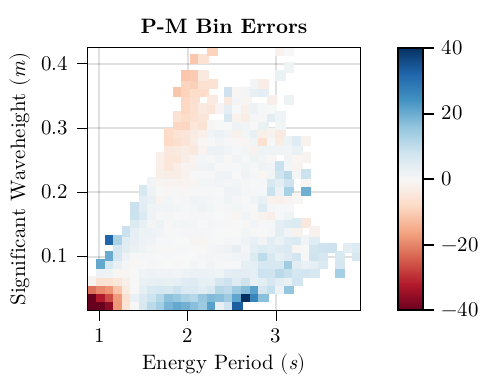}\label{fig:error_bins_a}} \\
        \subfigure[]{\includegraphics[width=0.95\linewidth]{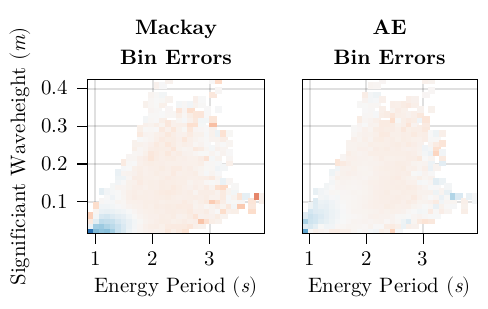}\label{fig:error_bins_b}}
        \caption{Per-bin error (\%) in power prediction. The largest errors for Pierson-Moskowitz, Mackay, and the autoencoder spectra are -45\%, 30\%, and 21\% respectively.}
        \label{fig:error_bins}
    \end{figure}

    Further investigation showed that the good performance of the Pierson-Moskowitz spectrum in the scaled sea state case is due to a symmetric distribution of large errors. 
    Figure \ref{fig:error_bins} shows the error for each $H_s$-$T_e$ bin for the $N=32$ case. 
    We can see in Figure~\ref{fig:error_bins_a}, that Pierson-Moskowitz has both large positive and negative errors.
    Pierson-Moskowitz's largest bin error is -45\% while the largest Mackay and the Autoencoder have is 30\% and 21\% respectively.
    Qualitatively, we can also see that both the Mackay and the autoencoder spectra have lower errors overall.
    
\begin{figure}[!ht]
        \centering
        \includegraphics[width=0.95\linewidth]{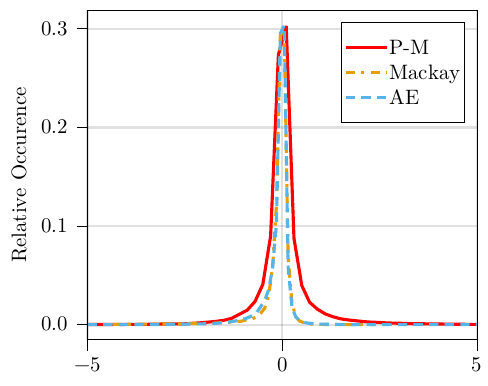}
        \caption{Distribution of the errors in power estimates for the scaled sea states case.}
        \label{fig:linear_relative_occurence}
    \end{figure}

\begin{figure}[!ht]
        \centering
        \includegraphics[width=0.95\linewidth]{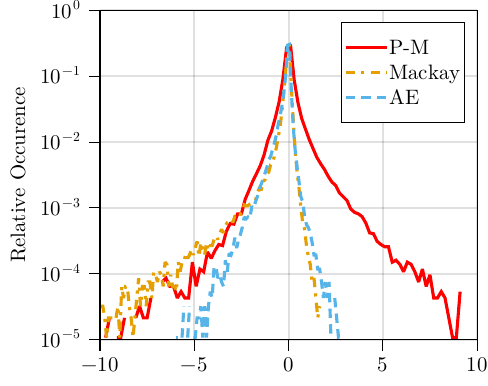}
        \caption{Distribution of the errors in power estimates for the scaled sea states case, in logarithmic scale.}
        \label{fig:log_relative_occurence}
    \end{figure}

    We can further confirm this by analyzing the distribution of all the errors (for each sample/spectra) shown in Figure~\ref{fig:linear_relative_occurence} and Figure~\ref{fig:log_relative_occurence} in linear and logarithmic scales respectively. 
    The logarithmic scale emphasizes the large error tails. 
    We can see that the error distribution of Pierson-Moskowitz is wider and symmetrical, on the other hand, both Mackay and the Autoencoder are sightly right skewed which explains their underestimations. 
    
\section{Conclusion}

    Two-parameter spectral models, which are ubiquitous in practice, are insufficient for accurately capturing the variance of spectra within bins\cite{SAULNIER2011130,MACKAY201617,merigaud2018power}.
    In this study we looked at the effects of this on the mean annual power estimates for WECs. 
    We considered two different 4-parameter spectra: one from the literature \cite{MACKAY201617} and our own site-specific model using an autoencoder. 
    These augment significant wave height and energy period with two additional \emph{shape parameters}.
    We compare the power estimate using these 4-parameter spectra to those using a two-parameter spectra. 
    Both approaches provide more accurate information leading to improved mean annual power predictions.
    In particular, the autoencoder performs the best achieving an error around 1\% in both case studies.
    
    The autoencoder provides a methodology for using buoy data to develop site-specific parameterization. 
    This can result in significantly better predictive performance at the cost of requiring training a new model for each site. 
    Transfer-learning could be used to significantly accelerate the training process for a new model by starting with the trained parameters of an existing model, such as those provided here. 
    In the rest of this section we discuss ways in which this methodology could be improved further. 

    Many aspects of the autoencoder presented here can likely be improved. 
    Different architectures and training hyperparameters could result in even better encoding. 
    The interpolation into a common set of normalized frequencies is inefficient since any given spectra contains energy in a small subset of these, and should be explored further. 
    The use of constant frequency spacing biases the discretization towards lower periods and the benefits of constant period spacing or other discretization approach should be explored. 
    Perhaps a method that allows using the frequencies as inputs, rather than assuming a predefined set of frequencies, could allow for using less inputs and outputs and parameterization approaches such as equal energy spacing.
    Finally, the physics informed layer could be improved to better deal with the discretization and interpolation errors. 

    The binning methodology for 4-parameter spectra also requires additional thought. 
    Standards such as the IEC provide best practice for binning significant wave height and  energy period. 
    Similar best practices could be developed for e.g. the Mackay spectra where the two shape parameters have physical meaning. 
    However, for the autoencoder the two parameters have no physical meaning and in fact every retraining of the autoencoder results in a different parameterization. 
    Methods for improving interpretability of these parameters and binning approaches should be explored. 

\section*{Acknowledgements}
    This research was supported by the U.S. Department of Energy’s Water Power Technologies Office.
    Sandia National Laboratories is a multi-mission laboratory managed and operated by National Technology \& Engineering Solutions of Sandia, LLC (NTESS), a wholly owned subsidiary of Honeywell International Inc., for the U.S. Department of Energy’s National Nuclear Security Administration (DOE/NNSA) under contract DE-NA0003525. 
    This written work is authored by an employee of NTESS. 
    The employee, not NTESS, owns the right, title and interest in and to the written work and is responsible for its contents. 
    Any subjective views or opinions that might be expressed in the written work do not necessarily represent the views of the U.S. Government. 
    The publisher acknowledges that the U.S. Government retains a non-exclusive, paid-up, irrevocable, world-wide license to publish or reproduce the published form of this written work or allow others to do so, for U.S. Government purposes. 
    The DOE will provide public access to results of federally sponsored research in accordance with the DOE Public Access Plan.

\def\url#1{}
\bibliographystyle{IEEEtran}
\bibliography{refs}

\appendix
\section{Linear Model of the WaveBot} \label{App:WaveBot}
    The WaveBot model is described in~\cite{michelen2023control} and summarized here. 
    Figure~\ref{fig:wavebot} depicts the model components and connections, and Table~\ref{tab:wavebot} summarizes all the model parameters. 
    For convenience, we express the model in terms of the radial frequency $\omega=2\pi f$. 

    \begin{table}[ht]
        \centering
        \caption{WaveBot model parameters.}
        \begin{tabular}{rc}
            \toprule
            \textbf{Parameter} & \textbf{Value} \\
            \midrule
            Mass, $M_h$ & 858 kg \\
            Hydrostatic stiffness, $K_{h}$ & 23.9E3 N/m \\
            Gear ratio, $N_d$ & 12 rad/m\\
            Drive-train inertia, $M_d$ & 2 kg\,m$^2$\\
            Drive-train friction, $B_d$ & 1 N\,m/(rad/s)\\
            Drive-train stiffness, $K_d$ & 0 N\,m/rad\\
            Torque constant, $K_\tau$ & 6.7 N\,m/A\\
            Winding resistance, $B_w$ & 0.5 $\Omega$\\
            Winding inductance, $L_w$ & 0 H \\
            \bottomrule
        \end{tabular}
        \label{tab:wavebot}
    \end{table}
    
    \subsection{Hydrodynamics}
        The intrinsic impedance $Z_i$ characterizes the hydrodynamic response of the hull to a wave with radial frequency $\omega$, and is given as 
        \begin{equation}
            Z_i(\omega) = (B(\omega) + B_h) + j\left(\omega(A(\omega)+M_h) - \frac{1}{\omega}K_h\right),
        \end{equation}
        where $j$ is the imaginary unit, $A(\omega)$ and $B(\omega)$ are the added mass and radiation damping coefficients obtained from solving the radiation problem using the boundary element method (BEM) code Capytaine~\cite{ancellin_capytaine_2019,babarit_theoretical_2015}, and $M_h$, $B_h$, and $K_h$ are the mass, hydrodynamic friction, and hydrostatic stiffness of the hull. 
        The velocity of the hull $\hat{U}$ when subject to wave excitation force $\hat{F}_e$ and PTO force $\hat{F}_p$ is 
        \begin{equation}
            \hat{U}(\omega) = \frac{\hat{F}_e(\omega) - \hat{F}_p(\omega)}{Z_i(\omega)}.
        \end{equation}
        The wave excitation force at a given frequency is given as 
        \begin{equation}
            \hat{F}_e(\omega) = \left(H_{FK}(\omega) + H_{diff}(\omega)\right) \hat{A}(\omega)\label{eq:excitation}
        \end{equation}
        where $H_{FK}$ and $H_{diff}$ are the Froude-Krylov and the diffraction coefficients, obtained from solving the diffraction problem using Capytaine, for a unit amplitude wave, and $\hat{A}$ is the complex wave amplitude. 
        The complex wave amplitude represents the sea state, as described in~\eqref{eq:wave_amplitudes}, and is the input to the model. 

    \subsection{Power Take-Off}
        The power take-off (PTO) is modelled as a two-port impedance that relates the flow variables velocity and current ($\hat{I}$) to the effort variables PTO force and voltage $\hat{V}$ as 
        \begin{equation}
            \begin{bmatrix} \hat{F}_p(\omega) \\ \hat{V}(\omega) \end{bmatrix} = 
            \begin{bmatrix} Z_{FU}(\omega) & Z_{FI}(\omega) \\ Z_{VU}(\omega) & Z_{VI}(\omega) \end{bmatrix} 
            \begin{bmatrix} \hat{U}(\omega) \\ \hat{I}(\omega) \end{bmatrix}
        \end{equation}
        \begin{align*}
            Z_{FU}(\omega) &= N_d^2Z_d(\omega) \\
            Z_{FI}(\omega) &= -\sqrt{\frac{3}{2}}K_\tau N_d \\
            Z_{VU}(\omega) &= \sqrt{\frac{3}{2}}K_\tau N_d \\
            Z_{VI}(\omega) &= Z_w(\omega),
        \end{align*}
        where the drive-train impedance $Z_d$ and generator's winding impedance $Z_w$ are given by 
        \begin{align}
            Z_d(\omega) &= B_d + j\left(\omega M_d -\frac{1}{\omega}K_d\right)\\
            Z_w(\omega) &=  B_w + j(\omega L_w)
        \end{align}
        and $N_d$, $M_D$, $B_d$, $K_d$ are the drive-trains's gear ratio, mass, friction, and stiffness, and $L_w$, $B_w$, and $K_\tau$ are the generator's winding inductance, winding resistance, and torque constant.

    \subsection{Equivalent Impedance and Optimal Controller}
        The model can be represented in terms of a single impedance, the Thévenin equivalent impedance $Z_{th}$ and voltage $\hat{V}_{th}$, given as
        \begin{align}
            \hat{V}_{th} &= \frac{Z_{VU}(\omega)}{Z_i(\omega) + Z_{FU}(\omega)}\hat{F}_e(\omega) \\
            Z_{th} &= Z_{VI}(\omega) - \frac{Z_{FI}(\omega)Z_{VU}(\omega)}{Z_i(\omega) + Z_{FU}(\omega)}.
        \end{align}
        The equivalent current is given then given as 
        \begin{equation}
            \hat{I}_{th}(\omega) = \frac{\hat{V}_{th}(\omega)}{2\mathcal{R}\left\{Z_{th}(\omega)\right\}},
        \end{equation}
        where $\mathcal{R}$ denotes the real part. 
    
        Based on impedance matching, the optimal controller consists of a load/controller impedance $Z_L=Z_{th}^*$, where a superscript $*$ indicates the complex conjugate. 
        Assuming an optimal controller the average electrical power is
        \begin{equation}
            P(\omega) = \frac{\left| \hat{V}_{th}(\omega) \right|^2}{8\mathcal{R}\left\{ Z_{th}(\omega) \right\}} . \label{eq:ipower}
        \end{equation}

    \subsection{Average Power at a given Sea State}
        For a sea state described by a discrete wave spectrum $S_i = S(\omega_i)$, the discrete wave amplitudes $\hat{A}_i = \hat{A}(\omega_i)$ are given as 
        \begin{equation}
            \hat{A}_i = \sqrt{2S_i \Delta \omega_i} \mathrm{e}^{j\phi_i}
            \label{eq:wave_amplitudes}
        \end{equation}
        where $\Delta \omega$ is the frequency spacing and $\phi$ are random phases.
        Since the spectra are reported at $N=47$ discrete frequencies, this results in $47$ discrete wave excitations components, using~\eqref{eq:excitation}, and discrete power components $P_i=P(\omega_i)$, using~\eqref{eq:ipower}. 
        The average power for the sea state is given as 
        \begin{equation}
            \overline{P} = \sum_{i=1}^{N} P_i .
        \end{equation}

    \subsection{Power Gain}
        The \emph{power gain} is the ratio of power delivered to the load, i.e.,electrical power produced, to the power available at the source. 
        In this case, since we are considering the optimal controller, the transducer power gain (based on actual power at the source) and the available power gain (based on maximum or ideal power at the source) are the same. 
        The power gain is given as~\cite{sandia_biconjugate}
        \begin{equation}
            G(\omega) = \left|\frac{Z_{VU}}{Z_{i}+Z_{FU}}\right|^2
            \frac{\mathcal{R}\left\{Z_i\right\}}{\mathcal{R}\left\{Z_{th}^*\right\}}
        \end{equation}
        where the frequency dependency of the various impedances was omitted for space. 

\end{document}

%% file: wavebot.pdf_tex
\begingroup%
  \makeatletter%
  \providecommand\color[2][]{%
    \errmessage{(Inkscape) Color is used for the text in Inkscape, but the package 'color.sty' is not loaded}%
    \renewcommand\color[2][]{}%
  }%
  \providecommand\transparent[1]{%
    \errmessage{(Inkscape) Transparency is used (non-zero) for the text in Inkscape, but the package 'transparent.sty' is not loaded}%
    \renewcommand\transparent[1]{}%
  }%
  \providecommand\rotatebox[2]{#2}%
  \newcommand*\fsize{\dimexpr\f@size pt\relax}%
  \newcommand*\lineheight[1]{\fontsize{\fsize}{#1\fsize}\selectfont}%
  \ifx\svgwidth\undefined%
    \setlength{\unitlength}{431.76961583bp}%
    \ifx\svgscale\undefined%
      \relax%
    \else%
      \setlength{\unitlength}{\unitlength * \real{\svgscale}}%
    \fi%
  \else%
    \setlength{\unitlength}{\svgwidth}%
  \fi%
  \global\let\svgwidth\undefined%
  \global\let\svgscale\undefined%
  \makeatother%
  \begin{picture}(1,0.38015529)%
    \lineheight{1}%
    \setlength\tabcolsep{0pt}%
    \put(0,0){\includegraphics[width=\unitlength,page=1]{wavebot.pdf}}%
    \put(0.44404673,0.27845928){\color[rgb]{0,0,0}\makebox(0,0)[lt]{\lineheight{1.25}\smash{\begin{tabular}[t]{l}\textit{$B_d$}\end{tabular}}}}%
    \put(0.46830307,0.1664523){\color[rgb]{0,0,0}\makebox(0,0)[lt]{\lineheight{1.25}\smash{\begin{tabular}[t]{l}\textit{$M_d$}\end{tabular}}}}%
    \put(0.3336274,0.11383193){\color[rgb]{0,0,0}\makebox(0,0)[lt]{\lineheight{1.25}\smash{\begin{tabular}[t]{l}\textit{Drive-train}\end{tabular}}}}%
    \put(0,0){\includegraphics[width=\unitlength,page=2]{wavebot.pdf}}%
    \put(0.57360522,0.20883208){\color[rgb]{0.10196078,0.10196078,0.10196078}\makebox(0,0)[lt]{\lineheight{1.25}\smash{\begin{tabular}[t]{l}\textit{$K_{\tau}$, $Z_w(\omega)$}\end{tabular}}}}%
    \put(0.8074624,0.22453902){\color[rgb]{0.10196078,0.10196078,0.10196078}\makebox(0,0)[lt]{\lineheight{1.25}\smash{\begin{tabular}[t]{l}\textit{$Z^*_{th}(\omega)$}\end{tabular}}}}%
    \put(0.57363302,0.24650845){\color[rgb]{0.10196078,0.10196078,0.10196078}\makebox(0,0)[lt]{\lineheight{1.25}\smash{\begin{tabular}[t]{l}\textit{Generator}\end{tabular}}}}%
    \put(0.82323522,0.10354203){\color[rgb]{0.10196078,0.10196078,0.10196078}\makebox(0,0)[lt]{\lineheight{1.25}\smash{\begin{tabular}[t]{l}\textit{Controller}\end{tabular}}}}%
    \put(0.16797497,0.05169186){\color[rgb]{0.10196078,0.10196078,0.10196078}\makebox(0,0)[lt]{\lineheight{1.25}\smash{\begin{tabular}[t]{l}\textit{Buoy}\end{tabular}}}}%
    \put(0.1677671,0.02162793){\color[rgb]{0.10196078,0.10196078,0.10196078}\makebox(0,0)[lt]{\lineheight{1.25}\smash{\begin{tabular}[t]{l}\textit{$Z_i(\omega)$}\end{tabular}}}}%
    \put(0,0){\includegraphics[width=\unitlength,page=3]{wavebot.pdf}}%
    \put(0.24876931,0.22453271){\color[rgb]{0,0,0}\makebox(0,0)[lt]{\lineheight{1.25}\smash{\begin{tabular}[t]{l}\textit{$N_d$}\end{tabular}}}}%
    \put(0,0){\includegraphics[width=\unitlength,page=4]{wavebot.pdf}}%
    \put(0.1795713,0.25981542){\color[rgb]{0,0,1}\makebox(0,0)[rt]{\lineheight{1.25}\smash{\begin{tabular}[t]{r}\textit{$\hat{U}(\omega)$}\end{tabular}}}}%
    \put(0,0){\includegraphics[width=\unitlength,page=5]{wavebot.pdf}}%
    \put(0.1795713,0.21769224){\color[rgb]{0,0,1}\makebox(0,0)[rt]{\lineheight{1.25}\smash{\begin{tabular}[t]{r}\textit{$\hat{F}_p(\omega)$}\end{tabular}}}}%
    \put(0.10443028,0.13113716){\color[rgb]{1,0,0}\makebox(0,0)[rt]{\lineheight{1.25}\smash{\begin{tabular}[t]{r}\textit{$\hat{F}_e(\omega)$}\end{tabular}}}}%
    \put(0.38503959,0.27782862){\makebox(0,0)[rt]{\lineheight{1.25}\smash{\begin{tabular}[t]{r}\textit{$K_d$}\end{tabular}}}}%
    \put(0.97925774,0.21895717){\color[rgb]{0,0,1}\makebox(0,0)[rt]{\lineheight{1.25}\smash{\begin{tabular}[t]{r}\textit{$\hat{V}(\omega)$}\end{tabular}}}}%
    \put(0.90798692,0.33617579){\color[rgb]{0,0,1}\makebox(0,0)[rt]{\lineheight{1.25}\smash{\begin{tabular}[t]{r}\textit{$\hat{I}(\omega)$}\end{tabular}}}}%
    \put(0.92233934,0.24913172){\color[rgb]{0,0,1}\makebox(0,0)[rt]{\lineheight{1.25}\smash{\begin{tabular}[t]{r}\textit{+}\end{tabular}}}}%
    \put(0.91625321,0.19725411){\color[rgb]{0,0,1}\makebox(0,0)[rt]{\lineheight{1.25}\smash{\begin{tabular}[t]{r}\textit{-}\end{tabular}}}}%
    \put(0,0){\includegraphics[width=\unitlength,page=6]{wavebot.pdf}}%
  \end{picture}%
\endgroup%